\documentclass{article}
\usepackage{amssymb}

\input{tcilatex}

\begin{document}

\subsection{\textbf{Direct observation of the effect of
isotope-induced-disorder on exciton binding energy in LiH}$_{x}$\textbf{D}$%
_{1-x}$\textbf{\ mixed crystals.}}

\bigskip

\ \ \ \ \ \ \ \ \ \ \ \ \ \ \ \ \ \ \ \ \ \ \ \ \ \ \ \ \textbf{\ \ \ \ \ \
\ \ \ \ \ \ \ \ \ \ \ \ \ \ \ \ \ \ V.G. Plekhanov}

\bigskip

\ \ \ \ \ \ \ \ \ \ \ \ \ \ \ \textbf{Computer Science College, Erika 7a,
Tallinn, 10416, ESTONIA }

\bigskip

\textbf{Abstract. }The results of the quantitative study of the
renormalization of binding energy the Wannier - Mott exciton by the isotope
effect are present for the first time. For this purpose accurate
measurements of the intrinsic luminescence and mirror reflection spectra of
LiH$_{x}$D$_{1-x}$ mixed crystals with a clean surface in the temperature
range 2 - 100 K were carried. Nonlinear dependence of exciton binding energy
E$_{b}$ on the isotope mass E$_{b}\sim $ f(x) is caused by the bands
fluctuation broadening which is connected with the isotope-induced-disorder.
Temperature dependence of exciton binding energy is briefly discussed. The
extrapolation of the asymptotic linear behavior of the exciton maximum
energy to T = 0 K enables to estimate the zero - point renormalization of
the exciton binding energy.

\bigskip\ \ 

\bigskip

PACS: 32.10B; 71.20P; 78.20; 78.40F.

\bigskip

A wide variety of novel isotope effects have been discovered on last four
decades [1-6] owing to the availability of high-quality bulk semiconductor
and insulator crystals with controlled isotopic composition (see, also,
reviews [7-10]). Recent high resolution spectroscopic studies of excitonic
and impurity transitions in high-quality samples of isotopically enriched Si
have discovered the broadening bound excitons emission (absorption) lines
connected with isotope-induced-disorder as well as the dependence of their
binding energy on the isotope mass [11-13]. The last effect was early
observed on the bound excitons in diamond [14-15] and more earlier on the
free excitons in LiH$_{x}$D$_{1-x}$ mixed crystals [16].

As is well-known (see, for example [16, 17, 4]) the band gap energy E$_{g}$
in the T$\longrightarrow $0 limit has a dependence on the average isotopic
mass $\overline{M\text{ }}$ due to two effects: a) the renormalization of E$%
_{g}$ by the electron-phonon interaction coupled with the dependence of the
zero-point amplitudes on $\overline{M}$ \ (see, also [18]) and b) the
dependence of the lattice constant on $\overline{M}$, leading to a change in
E$_{g}$ through the hydrostatic deformation potential. The electron-phonon
term is dominant [19, 20] \ and in the case of semiconductor crystals (C;
Ge; Si) with a weak isotope scattering potential is varied approximately as $%
\overline{M}^{-1/2}$. The value of the T =0 electron-phonon renormalization
energy contribution to E$_{g}$ can be independly determined from an
extrapolation of the high temperature linear dependence of E$_{g}\sim $ f(T)
to T=0 but again it should be remembered that E$_{g}\sim $ f(T) \ also has a
small contribution from change in volume with temperature. Therefore we
should distinguish between effects of the average atomic mass (which imply
the virtual crystal approximation - VCA) and effects of the mass
fluctuations (randomness) superimposed onto the virtual crystal.

In this communication we report the first results of the quantitative study
the dependence of free exciton's binding energy on isotope mass as well as
on the temperature in LiH$_{x}$D$_{1-x}$ mixed crystals. We'll show that the
exciton binding energy increase by 10$\pm $1 meV from $^{7}$LiH to $^{7}$%
LiD. Moreover, the dependence of E$_{b}$ $\sim $ f(x) is nonlinear. The last
effect, as will be shown below, is caused by the isotope-induced-disorder of
LiH$_{x}$D$_{1-x}$ mixed crystals. Briefly part of these results have been
published in [21].

Specimens of LiH, LiD LiH$_{x}$D$_{1-x}$ (0$\leq $ x $\leq $ 1) as well as
LiH$_{x}$F$_{1-x}$ were grown from melt using the modified Bridgman -
Stockbarger method (see, also [22]). This technique is described many times
early (see, e.g. [18, 2]). To improve the stoichiometric composition with
respect to hydrogen (deuterium), the crystals grown were additionally
annealed in an atmosphere of hydrogen or deuterium at a gas pressure of 3 to
5 atm. and a temperature 500 to 550$^{0}$C (melting point is 961 and 964 K
for LiH and LiD, respectively). For some crystals the time of heat treatment
was as long as 20 days (for more details see [2,7]).

Given the high reactivity of freshly cleaved LiH crystals \ in the
atmosphere, we had to develop a procedure of cleaving which would not only
ensure an initially clean surface, but also allow us to keep it fresh for a
few hours (the time taken to complete an experiment). These requirements are
satisfied by the well-known method of cleaving directly in the helium
chamber of the optical cryostat under liquid or superfluid helium, first
tried in Ref. [23]. We did not notice any changes in the spectra of
reflection or luminescence while working for 10 t0 16 hours with surface
prepared in this way. The device for cleaving the crystals had three degrees
of freedom and rotated through 90$^{0}$, which greatly helped in carrying
out the experiments. As a rule, specimens for experiments were cleaved off
bulk high - quality crystals. The experimental setup for measuring the low -
temperature reflection and luminescence spectra has been described more than
once (see, e.g. [7,18]) and consists of a double grating or prism
monochromators, an immersion helium cryostat, and a photovoltaic detector
(in photon counting mode). The results presented in this paper were obtained
from a clean crystal surface cleaved, as described above, directly in the
bath of helium cryostat. The spectra of mirror reflection were measured
using an the angle of incidence of 45$^{0}$. For our studies we selected
specimens which exhibited low dependence of exciton spectra of reflection
and luminescence on surface features.

As demonstrated earlier (see, e.g. [7]) most low-energy electron excitations
in LiH (LiD) crystals are large-radius excitons. The spectrum of exciton
photoluminescence of LiH crystals cleaved in liquid helium consists of a
narrow (in the best crystals, its half-width is the $\Delta $E $\leqslant $
10meV) phononless emission line and its broader phonon repetitions, which
arise due to radiative annihilation of excitons with the production of one
to five longitudinal (LO) phonons (see, Fig. 1). The phononless emission
line coincides \ in an almost resonant way with the reflection line (see,
insert Fig. 1) of the exciton ground state which is another indication of
direct electron transition (X$_{1}$ - X$_{4}$ [2]). The lines of phonon
replicas form an equidistant series biased toward lower energies from the
resonance emission line of excitons. The energy difference between these
lines is about 140 meV, which is close to the calculated energy of the LO
phonon in the middle of the Brillouin zone [21] and measured in [19]. \ The
isotopic shift of the zero-phonon emission line of LiH crystals equals that
in reflection spectra, 103 meV (see, also, insert in Fig.1). As the
deuterium concentrations increases, the width of the long-wave maximum in
reflection (and \ the width of the phononless line in luminescence) spectra
broadens and \ maximum shifts towards the shorter wavelengths. As can
clearly see in Fig. 1, all spectra exhibit a similar long-wave structure.
This circumstance allows us to attribute this structure to the excitation of
the ground (1s) and the first excited (2s) exciton states [2]. Three effects
are distinctly shown in the reflection spectra by an increase in deuterium
concentration:

1. The shortwavelength shift of the reflection (as luminescence) spectrum as
a whole;

2. The different velocity shift of the exciton maximum of the ground and the
first excited states on the temperature (see, Fig. 2);

3. Broadening of the long-wavelength maximum due to excitation of the ground
exciton state.

By the way we should note that the first excited exciton state is very
clearly seen in the luminescence spectra too (see Fig. 3). Although two
distinct contribution to $\Delta $E$_{g}$ (and respectively E$_{b}$) are
present \ that due to the effect of the vibration on the lattice constant
and the direct effect of the electron-phonon interaction [19, 20] for the
present task of the dependence of the exciton binding energy on the isotope
mass it will be considered only the mechanism of exciton-phonon interaction.

In accordance with the second effect in reflection spectra (see above point
2), as is to be expected, there is nonlinear dependence of the exciton
binding energy (E$_{b\text{ }}$ = 4/3$\Delta _{12}$) on the isotope
concentration. Displayed curve in Fig. 4 is the Wannier - Mott exciton
binding energy value which are determined by the hydrogenlike expression E$%
_{b}$ = $\frac{e^{4}\mu }{\text{2}\hslash ^{2}\varepsilon ^{2}\text{n}^{2}}$
using for this E$_{1}$ and E$_{2}$ values from the reflection and
luminescence spectra (see Figs. 1 and 3). The nonlinear nature of this
dependence is similar to the theoretical results [25] , where the influence
of the chemical disorder of the crystal lattice on the Wannier - Mott
exciton binding energy was seen and obtained a qualitative (not
quantitative!) agreement with experimental results on the GaAs$_{x}$P$_{1-x}$
[26] mixed crystals.

Before the comparison of our experimental results with the theory developed
by Elliott and Kanehisa [25], it would be prudent to briefly review main
properties of their theoretical model. One of the principal result of paper
[25] is the nonlinear dependence of exciton binding energy E$_{b}$ on the
concentration. As a consequence, the binding energy at half-and-half
concentrations is less than the value derived from the crystal virtual
model. According to Ref. [25] this model considers an exciton with a direct
gap of a semiconductor alloy. Such a system consists of an electron
(particle 1) in the conduction band (c) with mass m$_{\text{c}}$ and a hole
(particle 2) in the valence band (v) with mass m$_{\text{v}}$. The problem
of the exciton in disordered systems is to solve the Hamiltonian

H = $\overrightarrow{\text{p}}^{\text{2}}$/2m$_{\text{c}}$ +$\overrightarrow{%
\text{p}}^{\text{2}}$/2m$_{\text{v}}$+ u($\overrightarrow{\text{r}}_{\text{1}%
}$ - $\overrightarrow{\text{r}}_{\text{2}}$) + V$_{\text{c}}$($%
\overrightarrow{\text{r}}_{\text{1}}$) + V$_{v}$($\overrightarrow{\text{r}}_{%
\text{2}}$), \ \ \ \ (1)

with both the Coulomb interaction u and the potential V$_{v}$ due to
disorder ($\nu $ = c,v). Reference [25] neglected disorder - induced
interband mixing. As it is well known, in place of the electron-hole
coordinates, ($\overrightarrow{\text{r}}_{\text{1}}$, $\overrightarrow{\text{%
p}}_{\text{1}}$) and ($\overrightarrow{\text{r}}_{\text{2}}$, $%
\overrightarrow{\text{p}}_{\text{2}}$), one may introduce the center-of-mass
and relative coordinates, ($\overrightarrow{\text{R}}$, $\overrightarrow{%
\text{P}}$) and ($\overrightarrow{\text{r}}$, $\overrightarrow{\text{p}})$
to rewrite (1) as

H = $\overrightarrow{\text{p}}^{\text{2}}$/2$\mu _{\text{r}}$ + u($%
\overrightarrow{\text{r}}$) + $\overrightarrow{\text{P}}^{\text{2}}$/2M + V$%
_{\text{c}}$($\overrightarrow{\text{R}}$ + m$_{\text{v}}\overrightarrow{%
\text{r}}$/M) + V$_{\text{v}}$($\overrightarrow{\text{R}}$ - m$_{\text{c}}%
\overrightarrow{\text{r}}$/M), \ \ \ \ (2)

where $\mu _{r}$ and M are the reduced and total masses of excitons,
respectively. Because of the random potential, the translational and
relative degrees of freedom cannot be decoupled. This is essentially
difficult when considering the two-body problem in a disordered system (see
[25] and references therein). However, when the exciton state in question is
well separated from other states so that the energy spacing is much larger
than the translational width and disorder, one can forget about the relative
motion (H$_{r}$ = $\frac{\overrightarrow{p}^{2}}{\text{2m}_{r}}$ + u($%
\overrightarrow{r}$)) and just apply any single-particle alloy theory solely
to their translational motion. For each exciton state the translational part
of Hamiltonian in this case is

H$_{t}$ = $\overrightarrow{\text{P}}^{\text{2}}$/2M + \={V}$_{\text{c}}$($%
\overrightarrow{\text{R}}$) + \={V}$_{\text{v}}$(\={R}) \ \ \ \ (3).

Here \={V}$_{\text{c}}$ and \={V}$_{\text{v}}$ are averages of V$_{\text{c}}$
and V$_{\text{v}}$ with respect to the relative state $\phi $, for example:

$\overline{V_{c}}$($\overrightarrow{R}$) = $\dint $d$^{3}\overrightarrow{r}%
\left\vert \phi \left( \overrightarrow{r}\right) \right\vert $2V$_{c}$[$%
\overrightarrow{R}$ + $\frac{\text{m}_{\text{v}}}{\text{M}}\overrightarrow{r}
$] \ \ \ (4).

This approach is very similar to the Born-Oppenheimer adiabatic
approximation. Such situations hold in some mixed alkali halide crystals and
probably A$_{\text{2}}$B$_{\text{6}}$ crystals. On the contrary, when the
exciton binding energy is comparable to the disorder energy, the adiabatic
approximation breaks down, and it is essential to take into account the
effect of disorder on both the translational and relative motions. This is
the case with the Wannier-Mott exciton in A$_{\text{3}}$B$_{\text{5}}$
alloys, for which the Elliott and Kanehisa model was developed. In this case
the solution task is to start from the independent electron and hole by
neglecting u in (2) and then to take into consideration the Coulomb
interaction between the average electron and average hole. A further
simplified approach adopted in the literature (see, for example [27] and
references therein) in solving the Bethe-Salpeter equation [28] is to
suppose a free-electron-like one particle Green's function with a built-in
width to allow for the random potential due to disorder. \ In the cited
theoretical model [25], the average (or \textquotedblright virtual
crystal\textquotedblright ) gap is given by

E$_{\text{g}}^{\text{vc}}$(x) = E$_{\text{0}}$ + ($\delta _{\text{c}}$ - $%
\delta _{\text{v}}$)(x - 1/2), \ \ \ \ (5)

where E$_{\text{0}}$ is average gap, $\delta _{\text{c}}$, $\delta _{\text{v}%
}$ are the values of the fluctuation broadening of the conduction and
valence bands, respectively. Reference [25] also assumed the Hubbard density
of states for both the conduction and valence bands with width W$_{\text{c}}$
and W$_{\text{v}}$, respectively, as well as similar dispersion in both
bands. With this assumption the exciton binding energy has been calculated
according to the coherent potential approximation CPA model. \ As is
well-known, the main idea of the coherent potential methods in the
introduction of an auxiliary medium with a regular , i.e., spatially
periodic potential. By definition in the model this potential is also
complex. The formalism of the coherent potential model, convenient for
performing calculation and does not include fitting parameters, because of
using of the density of phonon states from virtual crystal model which used
virtual crystal approximation (VCA). It should be added here the key feature
of the model developed in Ref. 25 is the short-range nature of the Coulomb
potential (for details, see \ e.g. [7, 29]).

The data from Fig. 1 and other published sources \ [2, 7, 21] were used for
plotting the energy E$_{\text{b}}$ as a function of isotopic concentration x
in Fig. 4. The binding energy (defined as the band edge minus the exciton
energy) is given by [25]:

E$_{b}^{crys}$ = U$_{\text{0}}+\frac{\text{W}}{\text{2U}_{\text{0}}}$ + W \
\ \ \ \ \ \ (6).

In the last relation U$_{0}$ is the coupling constant at the total exciton
momentum $\overrightarrow{q}$ = 0.

Theoretical description of the binding energy of Wannier- Mott excitons as a
function of concentration x was based on the polynomial derived by Elliott
and cowokers [25]:

E$_{\text{b}}$ = E$_{\text{b}}^{\text{crys}}$- E$_{\text{bow}}\left[ \frac{%
\text{1-W}}{\text{2U}_{\text{0}}}\right] $ - E$_{\text{eff}}$, \ \ \ \ \ (7)

E$_{\text{eff}}$ = x $\left( \text{1-x}\right) $ $\frac{\delta _{\text{c}%
}\delta _{\text{v}}}{\text{W}}$, \ \ \ \ \ \ \ \ \ \ \ \ \ \ \ \ \ \ \ \ \ \
\ \ \ (8)

where W = W$_{\text{c}}$ + W$_{\text{v}}$, and W$_{\text{c}}$ and W$_{\text{v%
}}$ are the widths of the conduction band and the valence band which are
equal to 21 eV [30] and 6 eV [31, 32] respectively. Here E$_{\text{bow}}$ is
the curvature parameter found from the function E$_{\text{g}}$ $\propto $
f(x) (E$_{bow}$ = 0.046 eV [7]); $\delta _{\text{c}}$ and $\delta _{\text{v}%
} $ are the magnitudes of the fluctuation smearing of the valence band and
the conduction band edges , $\delta _{\text{c}}$ = 0.103 eV and $\delta _{%
\text{v}}$ = - 0.331 eV. As follows from Fig. 4, these values of the
parameters give a good enough description of the nonlinear dependence of the
binding energy of Wannier-Mott exciton in disordered medium isotope - mixed
crystals LiH$_{x}$D$_{1-x}$. This agreement between theory and experiment
once again proves the inherent consistency of the model proposed by Kanehisa
and Elliott, since the isotopic substitution affects the short-range part of
the interaction potential.

In this way, the nonlinear dependence of the binding energy of Wannier-Mott
exciton is caused by isotopic disordering of the crystal lattice. As is seen
from Fig. 4 the exciton binding energy decreasing (relative linear law (VCA)
- see dashed line in Fig. 4) in the vicinity of the middle meaning
concentration really calls out the fluctuated broadening of the edge of the
conduction and valence bands. In accordance with the theoretical model the
last reason gives rise to the reduced E$_{\text{g}}$ and there by the
shallowing of the exciton levels and, respectively, the reduction of E$_{%
\text{b}}$.

As follows from Fig. 1, the addition of deuterium leads not only to the
short-wave shift of the entire exciton structure (with different rates for
1s and 2s states), but also to a significant broadening of the long-wave
exciton reflection line. This line is broadened 1.5 - 3-fold upon transition
from pure LiH to pure LiD. The measure of broadening was the halfwidth of
the line measured in the standard way (see e.g. [33]) as the distance
between the maximum and the minimum in the dispersion gap of the reflection
spectrum, taken at half-height. The concentration dependence of the
halfwidth ($\Delta $E$^{\text{R}}$) of the long-wave band in the exciton
reflection spectrum at 2 K is shown in Fig. 5. Despite the large spread and
the very limited number of concentrations used, one immediately recognizes
the nonlinear growth of $\Delta $E$^{\text{R}}$ with decreasing x. A similar
concentration dependence of $\Delta E^{\text{R}}$ in the low-temperature
reflection spectra of solid solutions of semiconductor compounds A$_{\text{2}%
}$B$_{\text{6}}$ and A$_{\text{3}}$B$_{\text{5}}$ has been reported more
than once (see e.g. the review of Elliott and Ipatova [34] and references
therein). The observed broadening of exciton lines is caused by the
interaction of excitons with the potential of large-scale fluctuations in
the composition of the solid solution. Efros and colleagues (see e.g. [35])
used the Lifshitz method of optimal fluctuation [36] to express the formula
for the concentration dependence of the broadening of exciton reflection
lines:

$\Delta $E$^{\text{R}}$ = 0.5$\alpha $ $\left[ \frac{\text{x}\left( \text{1-x%
}\right) }{\text{Nr}_{\text{ex}}}\right] ^{\text{1/2}}$. \ \ \ \ \ \ \ \ \ \
\ \ \ \ \ \ \ \ \ (9)

where N - the concentration of sublattices nodes where the isotope
substitutes are placed, $\alpha $ = dE$_{\text{g}}$/ dx; r$_{\text{ex}}$ is
the exciton radius which varies from 47 \AA\ to 42 \AA\ upon transition from
LiH to LiD [2]. The results of calculation according to Eq. (9) are shown in
Fig. 5 by a full curve.

The experimental results lie much closer to this curve than to the straight
line plotted from the virtual crystal model. At the same time it is clear
that there is only qualitative agreement between theory and experiment at x $%
>$ 0.5. Nevertheless, even this qualitative analysis clearly points to the
nonlinear dependence of broadening on the concentration of isotopes, and
hence to the isotopic disordering. Since isotopic substitution only affects
the energy of optical phonon for the first time, and, as a consequence, the
constant of exciton-phonon interaction (in the first place, the Fr\"{o}hlich
interaction g$_{\text{F}}^{\text{2}}$), the nonlinearity of functions $%
\Delta $E$_{\text{b}}$ $\propto $ f(x), $\Delta $E$^{\text{R}}$ $\propto $
f(x) is mainly related to the nonlinear behavior of g$_{\text{F}}^{\text{2}%
}\propto $ f(x) . In this way, the experimental study of the concentration
dependence of the exciton-phonon interaction constant may throw light on the
nature and mechanism of the large-scale fluctuations of electron potential
in isotopically disordered crystals.

Returning to the results of Fig. 2, let us add that the different
temperature dependence of exciton peaks of n = 1s and 2s \ exciton states
leads to the temperature dependence of the binding energies of Wannier -
Mott excitons

E$_{b}$ $\sim $ f (T) \ \ \ \ \ \ \ \ \ \ \ \ \ \ \ \ \ \ \ \ \ \ \ \ \ \ \
\ \ \ \ \ \ \ \ \ \ \ \ \ \ \ \ \ \ \ \ \ \ \ \ \ (10).

This problem has not received any adequate treatment. More specifically, the
energy of the exciton binding E$_{b}$ in LiH \ \ \ crystals (as well as in
mixed crystals LiH$_{x}$F$_{1-x}$ \ (LiD$_{x}$F$_{\text{1-x}}$ )) decreases
with increasing temperature, whereas E$_{b}$ increases for excitons of the
green and yellow series in Cu$_{2}$O crystals [34]. A linear approximation
of the exciton binding energy in LiD$_{\text{0.995}}$F$_{\text{0.005}}$ (see
curve 3, Fig. 2) representing E$_{b}$ at T=0 K gives E$_{b}$(0) $\cong $ 55
meV. From this value we can see that renormalization of the binding energy
by the zero-point vibrations equals approximately $\simeq $ 10 \% from this
value, that is, on the other hand, composes only half of renormalized
exciton binding energy by isotope effect ($\simeq $ 10 meV, see above). It
is not excluded that the other part of renormalized exciton binding energy
is caused by exciton-polar phonon interaction.

In conclusion, the exciton luminescence and reflection spectra are used in a
quantitative study of the isotopic and temperature effects in LiH$_{x}$D$%
_{1-x}$ mixed crystals with a clean surface. It was shown that the
short-range character of the potential of a disordered crystal lattice with
isotope substitution is responsible for the broadening of the valence and
conduction bands. Nonlinear dependence of the exciton binding energy on the
isotope mass E$_{b}$ $\sim $ f(x) is due the isotope-induced-disorder of LiH$%
_{x}$D$_{1-x}$ mixed crystals. Temperature dependence of exciton binding
energy is briefly discussed. The extrapolation of the asymptotic linear
behavior of the exciton maximum energy to T = 0 K enables to estimate the
zero - point renormalization of the exciton binding energy.

\bigskip

\textbf{Acknowledgements.} I would like to express my deep thanks to Prof.
K. Wadler for improving my English.

\bigskip

\textbf{References.}

1. F.I. Kreingold, K.F. Lider, L.E. Solov'ev, Sov. Phys. JETP Let., \textbf{%
23} (1976) 624.

2. V.G. Plekhanov, T.A. Betenekova, V.A. Pustovarov, Sov. Phys. Solid State 
\textbf{18} (1976) 1422;

\ \ \ \ V.G. Plekhanov, Physics - Uspekhi \textbf{40} (1997) 553.

3. V.F. Agekyan, V.M. Asnin, A.M. Kryukov, Sov. Phys. Solid State \textbf{31}
(1989) 2082.

4. A.T. Collins, S.C. Lawson, G. Davies and H. Kanba, Phys. Rev. Lett. 
\textbf{65} (1990) 891.

5. T.R. Anthony, W.F. Banholzer, J.F. Fleisher, L.-H Wei, P.K. Kuo, Phys.
Rev. \textbf{B42} (1990) 1104.

6. H.D. Fuchs, C.H. Grein, C. Thomsen, M. Cardona, Phys. Rev. \textbf{B43}
(1991) 4835.

7. V.G. Plekhanov, Isotope Effects in Solid State Physics, Academic Press,
New York, 2001.

8. M. Cardona, Phys. Stat. Solidi (b) \textbf{220} (2000) 5; Phys. Stat.
Solidi (a) \textbf{188} (2001) 1209.

9. E.E. Haller, J. Appl. Phys. \textbf{77} (1995) 2857; Solid State Commun. 
\textbf{133} (2005) 693.

10. M.L.W. Thewalt, Solid State Commun. \textbf{133} (2005) 715.

11. D. Karaiskaj,M.L.W. Thewalt, T. Ruf, M. Cardona, M. Konuma, Solid State
Commun. \textbf{123} (2002) 87; M. Cardona and M.L.W. Thewalt, Rev. Mod.
Phys. \textbf{77}, (2005) 1173.

12. D. Karaiskaj, T.A. Meyer, M.L.W. Thewalt and M. Cardona, Phys. Rev. 
\textbf{B68} (2003) 121201 (R).

13. S. Tsoi, H. Alawadhi, X. Lu, J.W. Ager III, C.Y. Liao, H. Rieman, Phys.
Rev. \textbf{B70} (2004) 193201.

14. H. Kim, S. Rodriguez, T.R. Anthony, Solid State Commun. \textbf{102}
(1997) 861.

15. M. Cardona, Solid State Commun. \textbf{121} (2002) 7.

16. A.A. Klochikhin and V.G. Plekhanov, Sov. Phys. Solid State \textbf{22}
(1980) 342; \ V.G. Plekhanov, Fiz. Tverd. Tela \textbf{38} \textbf{\ }(1996)
1159 (in Russian).

17. F.I. Kreingold, Fiz. Tverd. Tela \textbf{20} (1978) 3138 (in Russian).

18. V.G. Plekhanov, Progress in Materials Science \textbf{51} (2006) 287.

19. V.G. Plekhanov, Phys. Solid State (St-Petersburg) \textbf{35} (1993)
1493; J. Nuclear Science and Technology \textbf{43} (2006) 375.

20. S. Zollner, M. Cardona and S. Gopalan, Phys. Rev. \textbf{B45} (1992)
3376.

21. V.G. Plekhanov, Progr. Solid State Chem. \textbf{29} \ (2001) 77.

22. O.I. Tytyunnik, V.I. Tyutyunnik, B.V. Shulgin, F.F. Gavrilov, and G.I.
Pilipenko, J. Crystal Growth, \textbf{68} (1984) 741.

23. V.G. Plekhanov, A.V. Emelyanenko, A.U. Grinfelds, Phys. Lett. \textbf{%
A101} (1984) 291.

24. J.L. Verble, J.L. Warren, J.L. Yarnell, Phys. Rev. \textbf{168} (1968)
980.

25. M.A. Kanehisa and R.J. Elliott, Phys. Rev. \textbf{B35} (1987) 2228.

26. R.J. Nelson, in Excitons, ed. by E.I. Rashba and M.D. Sturge
(North-Holland, Amsterdam, 1982), 319.

27. A.A. Klochikhin, Sov. Phys. Solid State \textbf{22} (1980) 1690.

28. H.A. Bethe, E. Salpiter, Quantum theory of one and two electron atoms,
AQcademic Press, New York, 1957.

29. R.J. Elliott, J.A. \ Krumhansl, P.L. Leath, Rev. Mod. Phys. \textbf{46}
(1974) 465.\ 

30. J. Kama and N. Kawakami, Phys. Lett. \textbf{A126} (1988) 348.

31. T.A. Betenekova, I.M. Shabanova, F.F. Gavrilov, Sov. Phys. Solid State 
\textbf{20} (1978) 820.

32. K. Ichikawa, N. Susuki, K. Tsutsumi, J. Phys. Soc. Japan \textbf{50}
(1981) 3650.

33. N.N. Ablyazov, A.G. Areshkin, V.G. Melekhin, L.G. Suslina and D.L.
Fedorov, Phys. Stat. Solidi (b) \textbf{135} (1986) 217.

34. R.J. Elliott and I.P. Ipatova (Eds), Optical Properties of Mixed
Crystals (North-Holland, Amsterdam, 1988).

35. A.L. Efros and M.E. Raikh, in [34] Chapter 5.

36. I.M. Lifshitz, Selected works, Science, Moscow, 1987 (in Russian).

37. T. Itoh, S.J. Narita, J. Phys. Soc. Japan \textbf{39} (1975) 132.

\textbf{\bigskip }

\textbf{Figure captions.}

1. Fig. 1. Luminescence spectra of free excitons at 2 K in LiH and LiD
crystals cleaved in liquid helium. In insert: mirror reflection spectra of
crystals. \ Curve 1: LiH; curve 2: LiH$_{x}$D$_{1-x}$ and curve 3: LiD.
Curve 4 is the light source   without crystals.

2. Fig. 2. Temperature dependence of the distance between the
long-wavelength peaks ($\Delta _{12}$) in specular reflection spectra of
pure and mixed crystals: 1 - LiH; 2 - LiD; 3 - LiD$_{0.995}$F$_{0.005}$.

3. Fig. 3. The reflection (1) and luminescence (2) spectra of LiD crystal at
2 K.

4. Fig. 4. Concentration dependence of the binding energy of a Wannier -
Mott exciton at 2 K in LiH$_{x}$D$_{1-x}$ mixed crystals: 1 - VCA
approximation  model; 2 - calculation according to equation (8);
experimental points indicated by triangles.

5. Fig. 5. Concentration dependence of the half-width of the  ground state
line of the exciton in the mirror reflection spectrum at 2 K. 1 - VCA
approximation  model; 2 - calculation according to equation (9);
experimental points are indicated by crosses.

\end{document}